\documentclass[10pt]{article}

\usepackage{graphics}

\usepackage{graphics}
\usepackage{dcolumn}
\usepackage{amsmath}
\usepackage{amssymb}
\usepackage{graphics}
\usepackage{bm}
\newcommand{\bea}{\begin{eqnarray}}
\newcommand{\eea}{\end{eqnarray}}
\newcommand{\be}{\begin{equation}}
\newcommand{\ee}{\end{equation}}
\newcommand{\f}[1]{\frac{1}{2}}
\def\be{\begin{eqnarray}}
\def\ee{\end{eqnarray}}
\def\bd{\begin{displaymath}}
\def\ed{\end{displaymath}}

\def\etal{{\em et al.} }

\def\NP{{\em Nucl. Phys. A }}
\def\PR{{\em Phys. Rev. C }}

\def\PRL{{\em Phys. Rev. Lett. }}
\def\PL{{\em Phys. Lett. B }}
\def\jpg{{\em J. Phys. G: Nucl. Part. Phys. }}
\def\EPJ{{\em Eur. Phys. J. A }}
\def\IJME{{\em Int. J. Mod. Phys. E  }}
\def\IJMA{{\em Int. J. Mod. Phys. A }}

\begin{document}
\title
{ Strange baryons, nuclear dripline and shrinkage : A Relativistic Mean Field study}
\maketitle 
\newcounter{saveeqn}
 
\centerline{\large Bipasha Bhowmick, Abhijit Bhattacharyya and G. Gangopadhyay}
\vskip 0.3cm
\centerline{\large Department of Physics, University of Calcutta}
\centerline{\large 92, Acharya Prafulla Chandra Road, Kolkata-700 009, India}
\centerline{email: ggphy@caluniv.ac.in}


\begin{abstract}
Neutron and proton driplines of single-$\Lambda$ and double-$\Lambda$ hypernuclei, 
$\Xi^{-}$ hypernuclei as well as normal nuclei
are studied within a relativistic mean field approach using an extended form
of the FSU Gold Lagrangian density. Hyperons are found to produce bound nuclei
beyond the normal nuclear driplines. 
Radii are found to decrease in hypernuclei near the driplines, in line with
observations in light $\Lambda$ hypernuclei near the stability valley, 
The inclusion of a $\Xi^{-}$ introduces a much larger change in radii 
than one or more $\Lambda$'s. 
\end{abstract}

\section{Introduction}

The production mechanisms, spectroscopy and decay
modes of hypernuclear states have been the subject of many
theoretical and experimental studies\cite{1a,1b,1c,1d}. 
Theoretical models used in studies of hypernuclei
extend from non-relativistic approaches based on One Boson Exchange models, to the relativistic mean field (RMF)
approximation and quark-meson coupling models. However,
the present knowledge of the $\Lambda$-N interaction in particular, and of
hypernuclear systems in general, is restricted to the valley of
$\beta$ stability.
In view of recent advances in producing light nuclei very 
close to the proton and neutron drip lines using radioactive ion 
beams, study of ground-state properties of these nuclei 
has assumed importance. These exotic nuclei provide a testing 
 ground of various theoretical models, which must explain 
the systematics of various properties over long chains of isotopes.
 Exotic nuclei on the neutron-rich side are especially important
 in nuclear astrophysics as they are expected to play an
important role in nucleosynthesis by neutron capture.
 Knowledge about their structure and properties would
help in the determination of astrophysical conditions for the
formation of neutron-rich stable isotopes.
It has been suggested\cite{1e} that a study of $\Lambda$ 
hypernuclei with a large neutron excess might also display 
interesting phenomena. 
On one hand such hypernuclei,
 corresponding to core nuclei which are unbound or weakly bound,
are themselves of considerable theoretical interest. On the other hand, one
can speculate on the possible role of neutron-rich $\Lambda$
hypernuclei in the process of nucleosynthesis. 
 
One expects that neutron-rich hypernuclei, where a $\Lambda$ hyperon exists in the neutron-excess
environment, may provide more exotic candidates than the ordinary neutron-rich nuclei,
because the $\Lambda$ hyperon acts as a nuclear `glue' in nuclei, as pointed out by Majling\cite{1e}.
In fact,
it was found that the core-nucleus $\alpha + d$ in $^{7}_{\Lambda}$Li shrinks 
owing to
the glue-like role of the $\Lambda$\cite{2}. Moreover, the added $\Lambda$ can often make a system bound even if
the core-nucleus is unbound. For example, $^{6}_{\Lambda}$He has a bound 
ground state\cite{3} though the normal nuclear core $^5$He is unstable against neutron 
emission. The
presence of hyperons in high-density nuclear medium significantly affects the maximal mass of
neutron stars because it softens the Equation of State (EOS). However, the baryon fraction
 in neutron stars is expected to depend on properties of hypernuclear potentials\cite{4a,4b}. Therefore,
the study of neutron-rich hypernuclei is expected to shed light on the
hypernuclear potentials in neutron-rich environment.

Several experimental attempts to produce such neutron-rich $\Lambda$ hypernuclei were carried out
by double-charge exchange reactions such as (stopped $K^{-}, \pi^{+}$)\cite{5,6} and $(\pi^{-}, K^{+})$\cite{7}
on stable nuclear targets. Further experiments 
are planned at J-PARC facilities\cite{8} to investigate
more exotic structures of the neutron-rich hypernuclei.
Various efforts have also been made on the theoretical side.
For example, the relativistic Hartree-Bogoliubov model was
applied to the description of $\Lambda$ hypernuclei with a large neutron excess\cite{9}.
Samanta \etal used a generalized mass formula to calculate the neutron and proton drip lines of normal
and lambda hypernuclei\cite{10}.
In the present work the RMF theory is used to study the neutron and proton driplines of
single-$\Lambda$, double-$\Lambda$, $\Xi^{-}$ hypernuclei and compare them with the normal nuclear 
driplines. The FSU Gold Lagrangian density, which was extended to include hyperons\cite{11}, 
has been used for this purpose.

\section{Method}

In the RMF formalism\cite{12},  
There are a number of standard Lagrangian densities as well as a number of 
different parametrizations. In the present work the FSU Gold Lagrangian density\cite{13}
 has been employed.
 While similar in spirit to most other forces, it contains 
two additional non-linear meson-meson interaction terms in the Lagrangian 
density, whose main virtue is a softening of both the EOS of symmetric matter 
and the symmetry energy. 
As a result, the new parametrization becomes more 
effective in reproducing a few nuclear collective modes\cite{13}, namely the breathing 
mode in $^{99}$Zr and $^{208}$Pb, and the isovector giant dipole resonance in
$^{208}$Pb.
The hyperon part introduced into the Lagrangian density is 
\begin{eqnarray}
\mathcal{L}_{h} = \bar{\psi}_{h}[i\gamma_{\mu}\partial^{\mu}-M_{h}+g_{\sigma h}\sigma+ 
g_{\sigma^{\ast}h}\sigma^{\ast}-g_{\omega h}\gamma^{\mu}\omega_{\mu}
-g_{\rho h}
\gamma^{\mu}\rho_{\mu}-g_{\phi h}\gamma^{\mu}\phi_{\mu}]{\psi}_{h} 
\end{eqnarray}
The explanation of different terms of the above Lagrangian density may be obtained from Bhowmick \etal\cite{11}

In the present work neutron and proton rich hypernuclei have been studied in a 
RMF formalism in coordinate 
space in the spherical approximation. It is likely that some nuclei in the chain
are deformed and also possible that the actual position of the drip line may 
vary slightly on inclusion of deformation.  Some recent results can be found in Refs. 19 and 20.
However, as our aim is to 
investigate possible differences in the locations of the drip lines on addition 
of a hyperon, we expect our results to be valid even if deformation is taken 
into account. A $\delta$ interaction of 
constant strength, i.e. $V = V_{0} \delta(\vec{r_{1}} - \vec{r_{2}} )$ has been 
adopted for pairing correlation. The matrix elements are calculated 
using the wave functions obtained for the single particle states. We choose 
$V_{0} = -350$ MeV for the strength of the delta-interaction. This value of the 
strength has been used by Bhattacharya \etal\cite{15} and also by Yadav \etal\cite{16} 
The valence shell has always been taken as the pairing space for both protons and neutrons.

The usual
BCS equations now contain contributions from the bound states as well as the resonant continuum. 
For the equations involving the 
resonant continuum readers are referred to Sandulescu \etal\cite{14}
We have also included the effect of the width of the positive energy levels.
 The positive energy resonance solutions are obtained using the scattering
approach.
  We have used the tagging approximation for odd mass 
nuclei. 
In the present work the equations have been solved using 
a grid of size $0.08$ fm and a box size of 20 fm following the works of Bhattacharya\etal\cite{15}  They had verified that the
 results are unaffected if a larger
box size is taken. According to their calculations an increase to $22$ fm changes the total energy
by less than $0.005\%$ and radius by less than $0.05\%$ in $^{64}$Ca. Besides,
Zhou \etal\cite{15a}, have also found that 
the rms radii are independent of the box size for both stable and the dripline nuclei for a box of size $>20$ fm.

\section{Results}
\begin{figure}[b]
\center
\resizebox{8cm}{!}{ 
\includegraphics{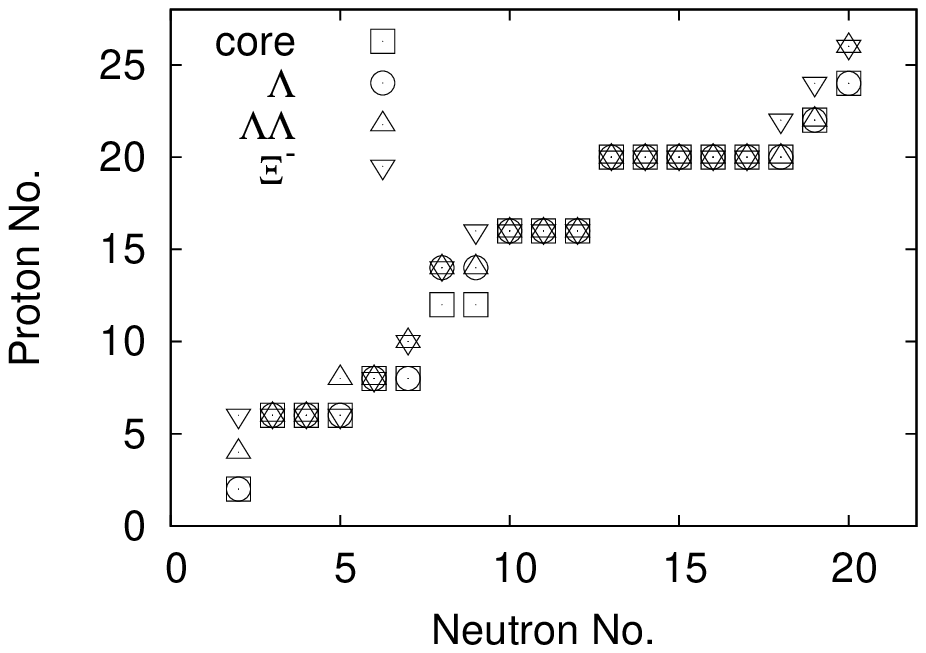}
}
\caption{
Proton driplines of single-$\Lambda$, double-$\Lambda$, $\Xi^{-}$ hypernuclei as compared to that 
of the normal nuclei for $N\le20$. }
\label{fig:1}  
\end{figure}

In Fig. 1 we present the results of our calculations for the proton driplines in
single-$\Lambda$, double-$\Lambda$ and $\Xi^{-}$ hypernuclei and compare them 
with the values obtained for the normal nuclei for $N\le20$. The bound nuclei, 
with the highest number of protons for a particular $N$, have been indicated.
The 
proton dripline does not change on addition of a single-$\Lambda$ hyperon to the normal
core, except for N=8 and 9. In these two instances the proton dripline shifts 
towards more proton rich side.
The inclusion of two $\Lambda$'s shifts the p-dripline away from the stability valley in nuclei with N=2, 5, 7, 8, 9 and 20. 
The inclusion of $\Xi^{-}$ has a more prominent effect on the proton dripline 
and shifts it even further for
N=2, 9, 18, 19 while being similar to double-$\Lambda$ in its effect on 
nuclei with $N=8$ and 20. Interestingly, at $N=5$ inclusion of two $\Lambda$'s,
and not $\Xi^-$, shifts the dripline.

\begin{table}[t]
\begin{center}
\caption{ Proton driplines for $\Lambda$, $\Lambda\Lambda$ and $\Xi^{-}$ hypernuclei compared to that of the 
normal nuclei for $20<N\le 82$.
}
\label{tab:1}       
\begin{tabular}{lclll|ccccc}\hline
N      &normal   &$\Lambda$    & $\Lambda\Lambda$  &$\Xi^{-}$  & N      &normal   &$\Lambda$    & $\Lambda\Lambda$  &$\Xi^{-}$\\
        &         &             &                 &            &        &          &           &                 &     \\\hline                                                     
27      &28       &28           &28               &30          &56      &52       &52          &52               &54    \\                
28      &30       &30           &30               &32          &59      &56       &56          &56               &58      \\                 
29      &32       &32           &32               &34          &66      &58       &58          &58               &60      \\  
32      &36       &36           &36               &38          &67      &58       &58          &58               &60  \\
33      &36       &38           &38               &38          &68      &60       &60          &60               &62   \\
34      &38       &38           &38               &40          &69      &60       &60          &60               &62   \\    
35      &38       &40           &40               &40          &70      &62       &62          &62               &64      \\
39      &40       &40           &40               &42          &73      &64       &64          &64               &66    \\
40      &40       &42           &42               &44          &75      &66       &66          &66               &68   \\
41      &42       &42           &42               &44          &77      &68       &68          &68               &70  \\
42      &44       &44           &44               &46          &79      &70       &70          &70               &72    \\
44      &46       &46           &46               &48          &81      &72       &72          &72               &74    \\
45      &48       &48           &48               &50          &82      &72       &72          &72               &74    \\
55      &50       &50           &50               &52          &        &         &            &                 &       \\\hline

\end{tabular}
\end{center}
\end{table}

In Table 1, we have noted the results of proton dripline study for $20<N\le 82$. 
For brevity, only those results have been tabulated where the dripline
shifts  due to the inclusion of hyperon(s). For example, there is no shift of
the dripline for $20<N<27$.
It is seen from Table 1 that the inclusion of a $\Lambda$ hyperon shifts the 
proton dripline towards more proton rich side in a few cases (N=33, 35, and 40).
The effect of two $\Lambda$'s is identical as the effect of inclusion of a 
 single-$\Lambda$.
The inclusion of $\Xi^{-}$ have much more pronounced effect
shifting the proton dripline in almost all the cases tabulated here except for N=33 and N=35.

From Fig. 1 and Table 1 we see
that the proton dripline of single $\Lambda$-hypernuclei 
shifts in a few light nuclei.
As the number of nucleon increases the dripline merges
with that for normal nuclei. This is a consequence of the weakness of the
$\Lambda$-nucleon coupling effect compared to the large nucleon-nucleon 
contribution to the static mean field potential.
In very light double-$\Lambda$ hypernuclei the effects are slightly more prominent than in single-$\Lambda$
hypernuclei. For higher masses the dripline merges with that of the normal nuclei as
in the case of single-$\Lambda$ hypernuclei.
The inclusion of a $\Xi^{-}$ hyperon affects the proton-driplines much more 
prominently than both of the single and two
$\Lambda$ hyperons.
It can be seen from Fig. 1 that the effect of inclusion of hyperons on the proton dripline
is similar irrespective of whether N is odd or even. 
Also we see that at certain places, the p-dripline does not change at all; this may be attributed to shell 
effects. When the drip line corresponds to a filled proton or neutron shell, in some
cases the excess 
binding energy from addition of hyperon(s) perhaps may not play an important 
role in shifting the dripline. As one can see in Fig. 1, $Z=8, 16$ and 20 
correspond to such positions. Near drip line, $N=16$ may act as a magic number. A similar situation is observed in 
$N=50$ and 58, the last being a subshell closure.

\begin{figure}[t]
\center
\resizebox{8cm}{!}{ 
\includegraphics{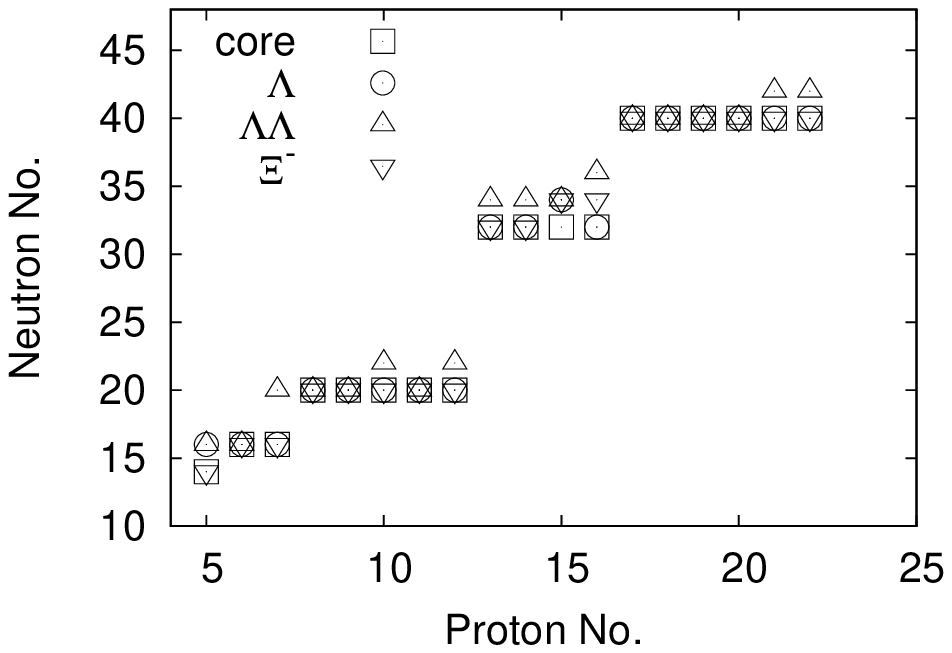}
}
\caption{
Neutron driplines of single-$\Lambda$, double-$\Lambda$, $\Xi^{-}$ hypernuclei as compared to that 
of the normal nuclei for $Z\le22$. }
\label{fig:2}  
\end{figure}
In Fig. 2 we present the results of our calculations for the neutron driplines
(for even neutron number only)
 in hypernuclei and compare them with that
obtained for the normal nuclei for $Z\le22$. We see that in all the cases except for Z=5 and
 Z=15 the neutron dripline does not change due to inclusion of a 
single-$\Lambda$ hyperon. The inclusion of two $\Lambda$'s 
 shifts the dripline towards more neutron rich side for certain
proton numbers, the effect being maximum for Z=7 and Z=16.
The effect of inclusion of $\Xi^{-}$ is manifested only for Z=15 and 16.
We see that at Z=15, the effect of all the hyperons are visible in the shifting of neutron dripline.

\begin{table}
\begin{center}
\caption{Neutron driplines for $\Lambda$, $\Lambda\Lambda$ and $\Xi^{-}$ hypernuclei compared to that of the 
normal nuclei for $22<Z\le50$.
}
\label{tab:2}       
\begin{tabular}{|c|c|c|c|c|}\hline
Z      &normal  &$\Lambda$  & $\Lambda\Lambda$  &$\Xi^{-}$  \\
             &1n(2n)            &1n(2n)              &1n(2n)                       &1n(2n)            \\\hline
26           &49(56)            &49(56)              &57(58)                       &49(56)             \\
27           &49(56)            &49(56)              &57(58)                       &49(56)             \\
28           &55(56)            &49(56)              &57(58)                       &49(56)              \\
29           &55(56)            &49(56)              &57(58)                       &49(56)             \\
30           &55(58)            &67(68)              &67(68)                       &49(68)            \\
31           &63(64)            &67(68)              &67(68)                       &67(68)                 \\
35           &79(80)            &79(80)              &79(82)                       &79(80)                \\
37           &79(80)            &79(80)              &83(84)                       &79(86)             \\
38           &79(86)            &81(86)              &83(88)                       &79(86)               \\
39           &79(86)            &81(86)              &83(88)                       &79(86)                \\
41           &81(86)            &81(86)              &83(92)                       &81(90)               \\
42           &81(90)            &81(90)              &83(92)                       &81(108)              \\
43           &81(90)            &81(90)              &83(92)                       &81(112)              \\
44           &81(90)            &81(90)              &83(92)                       &81(112)              \\
45           &81(90)            &81(90)              &83(92)                       &81(112)              \\
46           &81(90)            &81(90)              &83(92)                       &81(112)              \\
47           &81(90)            &81(90)              &83(92)                       &81(112)              \\
48           &81(110)           &81(110)              &83(112)                     &81(112)              \\
49           &81(110)           &81(110)              &83(112)                     &81(112)              \\
50           &109(110)          &109(110)             &111(114)                    &111(112)                  \\\hline

\end{tabular}
\end{center}
\end{table}

In Table 2, we have noted the results of neutron dripline study for $22<Z\le 50$. 
One important point that should be noted is that in heavier nuclei, the neutron drip line may
differ substantially for even and odd neutron numbers due to the effect of 
pairing.
Only those results have been tabulated for which there is a change in the 
dripline due to the inclusion of $\Lambda$, $\Lambda\Lambda$, or $\Xi^{-}$. 
The dripline shifts towards more neutron rich side in case of Z= 29, 30, 31, 38 
when a $\Lambda$ is present.
 In the neutron dripline side the inclusion of two $\Lambda$'s is really effective in shifting the dripline and its effect 
shows in all 
the cases presented in Table 2 in the shifting
of neutron drip-line. 
However the effect of $\Xi^{-}$ on the neutron dripline is much less visible. 
The inclusion of $\Lambda$ binds the neutron 
single particle levels strongly, while leaving the proton single particle levels 
unchanged, which may be a possible reason for this. 
From Fig. 2 and Table 2 we see that the effect of inclusion of strange baryons, on the neutron dripline
seems to be similar irrespective of whether Z is odd or even. 
Shell effects are also visible at certain neutron and proton numbers.

It is noticeable that the results of Ref.\cite{16a} for the Ca isotopes differs somewhat from our results.
In that work, it was predicted that the dripline nucleus for hypernuclei of Ca is at N=54 compared with that
 for ordinary nuclei at N=52. However we see no shift of dripline for the Ca nucleus.

We have also investigated the change in the root mean square (r.m.s) charge radii and baryon radii due to the
inclusion of a hyperon near the drip line. It is seen, that the shrinkage phenomenon in the light-mass
$\Lambda$ hypernuclei is present at the dripline and this effect enhances if an additional
$\Lambda$ is included. However there is a characteristic difference in the 
effect near the proton and the neutron driplines. 
At the p-dripline the r.m.s charge radius tends to decrease for $N\le4$, whereas at the n-dripline it is  the neutron
r.m.s radius that decreases for $Z\le7$ and also for the hypernucleus 
$^{35}_\Lambda$Si. 
However, these
changes in radii are not quantitatively much significant (below $1\%$).

On the other hand, the inclusion of a $\Xi^{-}$ may change the radii 
significantly. In Figs. 3 and 4 we plot the change in mean square radii 
$\Delta r^{2} = \langle r^{2} _{H} \rangle - \langle r^{2} _{core} \rangle$ at the proton and the neutron 
driplines respectively for $\Xi^{-}$ hypernuclei. Here, the suffices $H$ and 
$core$ refer to hypernucleus and normal core, respectively. 

\begin{figure}[h]
\center
\resizebox{9.5cm}{!}{ 
\includegraphics{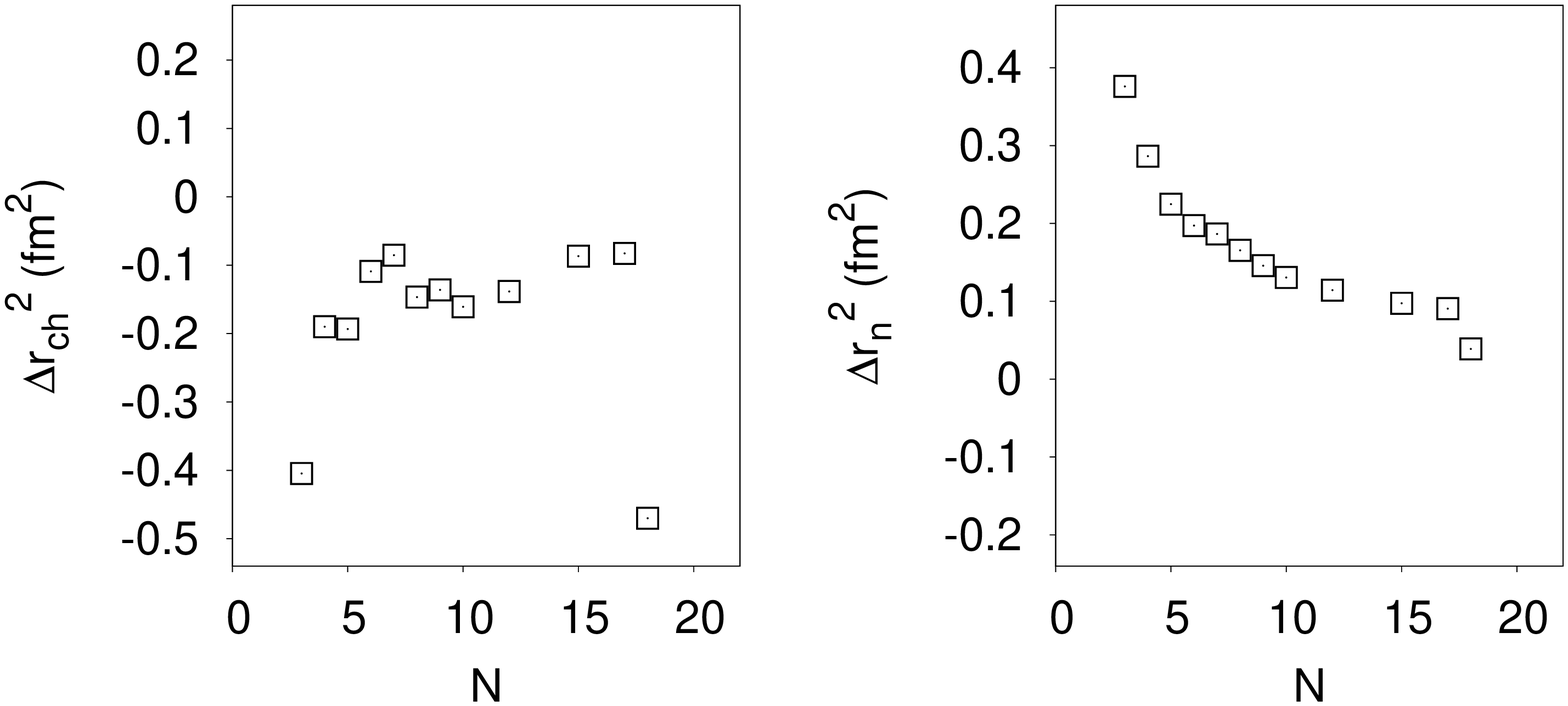}
}
\caption{
Change in mean square radius due to the inclusion of $\Xi^{-}$ at or near the proton driplines for  $N\le20$.
see text for details.}
\label{fig:3}  
\end{figure}

\begin{figure}[htb]
\center
\resizebox{9.5cm}{!}{ 
\includegraphics{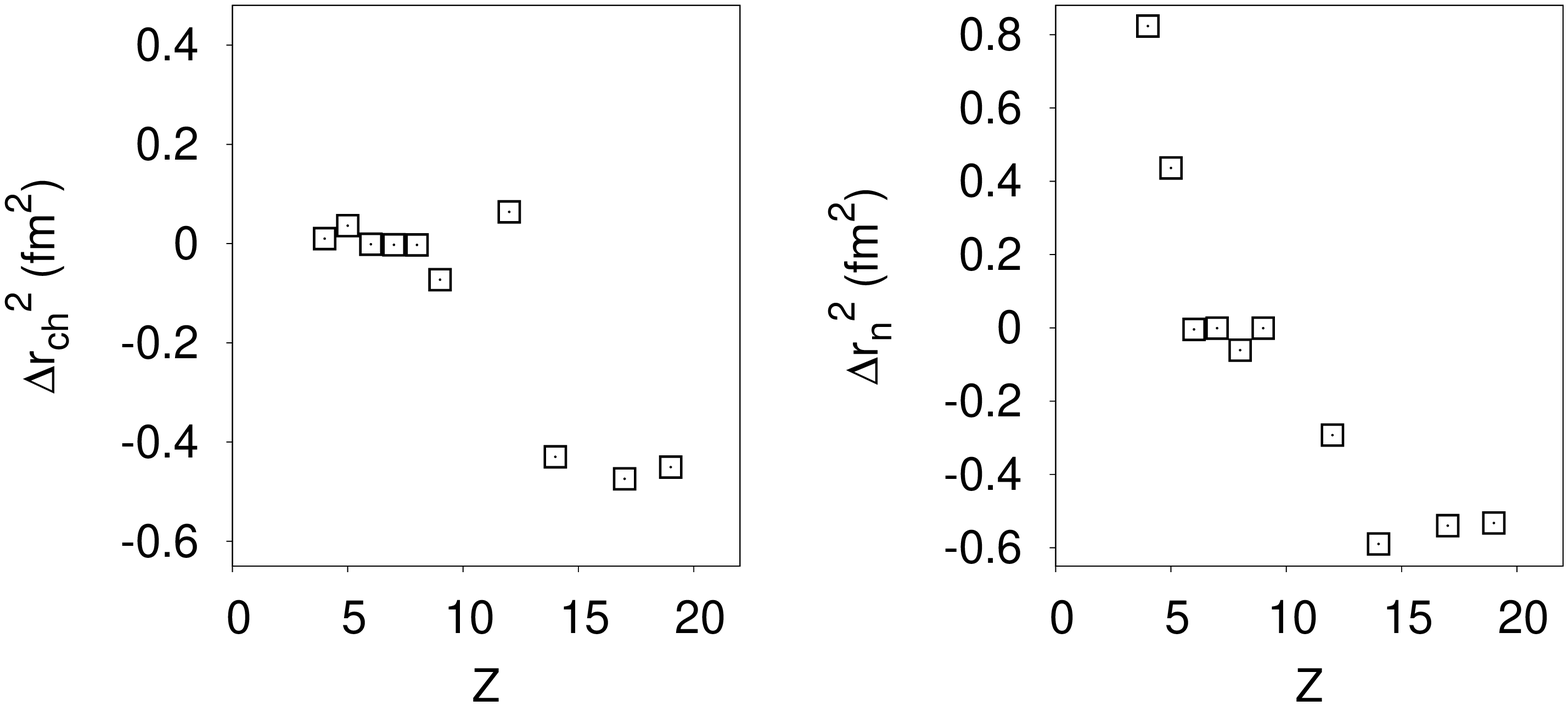}
}
\caption{
Change in mean square radius due to the inclusion of $\Xi^{-}$ at or near the neutron driplines for  $Z\le20$.
see text for details.}
\label{fig:4}  
\end{figure}

From Fig. 3 it can be seen that at the p-dripline of $\Xi^{-}$ hypernuclei the r.m.s charge 
radius decreases while the neutron radius
increases. A similar effect was seen in Tan \etal\cite{17} for nuclei away from dripline. There,
 a slow decrease in this effect with mass was seen. From the figure it is clear that
this effect seems to die out with mass number.

Fig. 4 shows that near the neutron dripline
in addition to the shrinkage in the charge radius, the neutron r.m.s radius also tends to shrink in some nuclei. The shrinkage effect increases as we go 
towards higher mass number and seems to saturate eventually.

These changes can be attributed to the effect of the charge and isospin of the 
$\Xi^{-}$. The changes in the scalar and the vector fields due to 
inclusion of a $\Xi^{-}$ are almost identical, so that these changes cancel out 
each other. On the other hand, both the isovector and photon 
fields change significantly. The isovector field,  being governed by the 
total isospin projection value, has opposite signs at the two driplines.  
This should increase the proton radius and decrease the neutron radius at the 
n-dripline. However the photon field,  which is more dominant,
compensates for the change in proton radii, so that both the proton and neutron 
radii decrease at the neutron dripline. This in agreement with the work by Tan 
\etal\cite{17} where, on switching off the coupling of the $\Xi^{-}$ to the $\rho$-meson, only the proton radius 
decreased leaving the neutron radius unchanged. As the shrinkage due to the $\Xi^{-}$ can 
be seen even at very low mass, it will be interesting to look for such changes experimentally.

\section{Summary}

To summarize, the effect of the addition of a hyperon to a non-strange
normal nucleus appears significant on both the neutron and the proton
driplines, particularly in lighter mass regions. The presence of hyperons creates bound states beyond the normal
dripline in many cases.
Inclusion of two $\Lambda$'s changes the neutron-dripline in many cases 
whereas the effect of $\Xi^{-}$ on the proton dripline is more prominent.
The latter effect may be attributed to the negative charge of the $\Xi^{-}$. 
It is seen that the inclusion of one or more $\Lambda$ makes the neutron 
single particle levels substantially more bound than the respective core 
nucleus while leaving the proton single particle levels unchanged, which 
may be the possible reason for the former effect. 
The shrinkage in size, seen 
in light-mass $\Lambda$ hypernuclei, is present at or near the dripline also. 
However the effect is different at the proton and neutron driplines. 
The inclusion of a $\Xi^{-}$ decreases the radius more effectively than a 
$\Lambda$ hyperon. 

\section*{Acknowledgment}

This work was carried out with financial assistance of the UGC (RFSMS, DRS, 
UPE) and DST, Government of India.

\end{document}